\documentclass[final,5p,times,twocolumn]{elsarticle}
\usepackage{bm}
\usepackage{amssymb}
\usepackage[colorlinks=true,linkcolor=black, citecolor=blue, urlcolor=blue]{hyperref}
\usepackage{amsthm}
\usepackage[fleqn]{amsmath}
\usepackage{float}
\usepackage{subfigure}
\usepackage{caption}
\usepackage{upgreek}
\usepackage{multirow}
\usepackage{array}
\biboptions{numbers,sort&compress}
\journal{Materials Science and Engineering: A}
\begin{document}
\captionsetup[figure]{labelfont={bf},labelformat={default},labelsep=period,name={Fig.}}
\captionsetup[table]{labelfont={bf},labelformat={default},labelsep=period,name={Table}}
\begin{frontmatter}

\title{Properties of dislocation lines in crystals with strong atomic-scale disorder}

\author[label1]{Jian-Hui Zhai}
\author[label1,label2]{Michael Zaiser\corref{cor1}}
\cortext[cor1]{Corresponding Author.}
\ead{michael.zaiser@ww.uni-erlangen.de}
\address[label1]{Institute of Materials Simulation
	(WW8), Friedrich-Alexander
	University Erlangen-N{\"u}rnberg
	(FAU), Dr.-Mack-Str. 77, 90762
	F{\"u}rth, Germany}
\address[label2]{Department of Engineering and Mechanics, Southwest Jiaotong University, Chengdu,
	 People$'$s Republic of China}

\begin{abstract}
We use a discrete dislocation dynamics (DDD) approach to study the motion of a dislocation under strong stochastic forces that may cause bending and roughening of the dislocation line on scales that are comparable to the dislocation core radius. In such 
situations, which may be relevant in high entropy alloys (HEA) exhibiting strong atomic scale disorder, standard scaling arguments based upon a line tension approximation may be no longer adequate and corrections to scaling need to be considered. We first study the wandering of the dislocation under thermal Langevin forces. This leads to a linear stochastic differential equation which can be exactly solved. From the Fourier modes of the thermalized dislocation line we can directly deduce the scale dependent effective line tension. We then use this information to investigate the wandering of a dislocation in a crystal with spatial, time-independent ('quenched') disorder. We establish the pinning length and show how this length can be used as a predictor of the flow stress. Implications for the determination of flow stresses in HEA from molecular dynamics simulations are discussed. 
\begin{flushright}

\end{flushright}
\end{abstract}


\end{frontmatter}


\section{Introduction}
\label{}
Discrete dislocation dynamics (DDD) as a mesoscale method to simulate plastic deformation by considering dislocation motion, reactions, and interactions with other defects has been developed over the past decades to simulate deformation of bulk materials \cite{Zbib1998_IJMS,Verdier1998_MSMSE} and micropillars \cite{Schwarz1999_JAP,Ghoniem1999_PRB,Weygand2002_MSMSE}. Most of these simulations consider the evolution of dislocation systems in a deterministic setting, without explicitly accounting for stochastic influences that may be either due to thermal fluctuations ('annealed disorder') or due to spatially random but temporally constant forces arising from small-scale (e.g. chemical) disorder ('quenched disorder'). Such random effects may be significant in novel materials with a large degree of atomic scale disorder such as HEA. They are the main focus of the present study.

Thermal effects can be considered by incorporating Langevin forces into the equations of motion \cite{Ronnpagel1993_PSS,Mohles1996_CMS,Hiratani2002_JEMT,Hiratani2003_JNM}. R{\"o}nnpagel et al. \cite{Ronnpagel1993_PSS} developed a stochastic model that considers the effects of temperature in a line tension model within Brownian dynamics scheme. They simulated dislocation glide in a stress field which was generated by localized obstacles and found that the effective attack frequency (inverse of waiting time) was indepedent of the choice of segment length and the activation enthalpy was dependent on applied stress and temperature. Mohles and R{\"o}nnpagel \cite{Ronnpagel1993_PSS} studied dislocations in a field of obstacles or interacting with obstacle walls. They found that the activation volume for overcoming such obstacles was independent on the spacing of obstacle and the drag coefficient had no influence on waiting time. Their simulation system focused on two dimensional problems \cite{Ronnpagel1993_PSS,Mohles1996_CMS}. Hiratani and Zbib \cite{Hiratani2002_JEMT,Hiratani2003_JNM} extended  the model proposed by R{\"o}nnpagel et al. \cite{Ronnpagel1993_PSS} to three-dimensional simulations and used it to simulate dislocation glide through weak obstacles which were represented by stacking fault tetrahedra (SFTs). They found that dislocation motion was obstacle-controlled when the applied stress was below a critical resolved shear stress (CRSS), otherwise drag-controlled \cite{Hiratani2002_JEMT}; the dislocation line was found to exhibit a self-affine structure as manifested by non-trivial height-difference correlations of dislocation shapes \cite{Hiratani2003_JNM}. Langevin forces were implemented in different ways in these simulation schemes. Some researchers directly applied Langevin forces on dislocation nodes \cite{Ronnpagel1993_PSS,Mohles1996_CMS}, while others applied Langevin force on dislocation segments \cite{Hiratani2002_JEMT,Hiratani2003_JNM}. In this paper we show that, to prevent artefacts, in a nodal dislocation dynamics scheme Langevin forces should be directly applied to dislocation nodes. 

Solid solution strengthening is a common method to strengthen materials where foreign atoms interact with dislocations and impede their movement \cite{Argon2008_OUP} ('pinning'). This is one particular example of a class of problems studied extensively in statistical physics, namely the pinning of elastic manifolds by random fields, see \cite{Chauve2000_PRB}. Concepts of elastic manifold depinning and the associated statistical phenomena were applied to the athermal motion of dislocations by Zapperi and Zaiser \cite{Zapperi2001_MSEA} and Bako et. al. \cite{Bako2008_PRB}, and to dislocation motion at finite temperature by Ioffe and Vinokur \cite{Ioffe1987_JPC} and Zaiser \cite{Zaiser2002_PM}. 

The behavior of dislocations in pinning fields created by the superposition of forces from multiple pinning centers depends crucially on the minimal wavelength of dislocation shape fluctuations. If this wavelength is comparable to the spacing of pinning centers along the dislocations, i.e., if the dislocation bends around the pinning centers individually, we speak of strong pinning, otherwise if the wavelength is larger than the pinning center spacing, we speak of weak or collective pinning.  Fleischer \cite{Fleischer1961_AM,Fleischer1963_AM} considered solute atoms as strong pinning centers. Labusch \cite{Labusch1970_PSS,Labusch1972_AM} developed a statistical model to consider describe the interaction between dislocations and solute atoms where the solute atoms act as weak pinning centres atoms. The Fleischer model is appropriate to dilute solid solutions, where spacings between solute atoms are large, whereas the Labusch model is suitable for concentrated solid solutions and has been also applied to high entropy alloys (HEAs) \cite{Toda2015_AM,Wu2016_AM,Varvenne2016_AM,Varvenne2017_AM}. Note that the concepts of 'strong' vs 'weak' pinning do not relate to the magnitude of the CRSS: In highly dilute solid solutions, the CRSS is low but pinning may be 'strong' in the above defined technical sense, whereas in HEA the CRSS is high but pinning may be technically 'weak'.  

Theoretical approaches to dislocation pinning, envisaged as elastic manifold pinning, often rely on a line tension approximation. This idea may be problematic in HEA where the atomic scale disorder may be strong and cause roughening of dislocations down to the nanometre scale, such that the minimal dislocation 'wavelength' is no longer small compared to the dislocation core radius. Here we investigate systematically the corrections that need to be applied to standard pinning theories as a consequence. The remainder of this paper is organized as following. Section \ref{sec:2} gives an overview of results from statistical physics concering the dynamics of elastic lines under the influence of random forces, and introduces several concepts that will be used in Sections to analyze the results of the dislocation dynamics simulations.

Section \ref{sec:3} introduces discrete dislocation dynamics (DDD) under the influence of Langevin forces in \ref{sec:3.1}, the model is validated in \ref{sec:3.2}. Results of simulations are reported in Section \ref{sec:4}, where spatio-temporal roughening under the influence of thermal forces is studied in \ref{sec:4.1}. Comparison with analytical predictions from Section \ref{sec:2} allows us to determine an effective, scale dependent line energy which is used in Section \ref{sec:4.2} for studying dislocation glide under the influence of spatially random forces and  applied stresses. Relations are given that allow to determine a characteristic pinning length from the zero-stress relaxed structure of the dislocation, and to relate this pinning length to the zero-temperature CRSS. Implications for determining CRSS values in HEA from molecular and ab-initio simulations are discussed and conclusions are given in Section \ref{sec:5}.

\section{Scaling theory of elastic lines in random fields}
\label{sec:2} 

\subsection{Thermal fluctuations}

The thermal equilibrium shape of an elastic line under the influence of thermal fluctuations can be deduced from simple thermodynamic arguments. We consider an elastic line of length $L$ and line energy ${\cal T}$ pinned at both ends. The initial line direction is identified with the $x$ direction of a Cartesian coordinate system. The fundamental modes of the line are given by $y_n(x) = A_n \sin(n \pi x/L)$ where $n$ is a positive integer number. The associated energy is 
\begin{equation}
E_n = \frac{{\cal T}}{2} \int \left(\frac{\partial y_n(x)}{\partial x} \right)^2 dx = n^2 \frac{{\cal T} \pi^2}{4 L} A_n^2.
\end{equation}
From the equipartition theorem it then follows that $E_n = k_BT/2$, hence 
\begin{equation}
A_n^2 = \frac{2k_B T}{ {\cal T}L} \left(\frac{L}{\pi n}\right)^2,
\label{eq:intens}
\end{equation}
i.e., the square mode amplitude ('intensity') is expected to be proportional to temperature and inversely proportional to the square of the wave number. Conversely, from measurements of the mode amplitude the line energy ${\cal T}$ can be deduced as 
\begin{equation}
{\cal T} =  \frac{2 k_B T}{L} \frac{1}{(\pi n A_n)^{2}}
\label{eq:linetens}
\end{equation}
For the case of dislocations, this provides an easy check to see whether a line tension approximation is warranted or whether corrections must be taken into account.

Defining the mode wave vector as $q = n \pi/L$ and noting that the phases of the different Fourier modes are independent random variables, we can deduce, in the limit $L \to \infty$ the scaling of the power spectrum of the line as $P(k) \propto q^{-a}$ where $a = 2$. Using general results for self affine manifolds \cite{Schmittbuhl1995_PRE}, this implies that the line shape represents, in the limit $L \to \infty$, a self affine fractal with roughness exponent $\zeta = (a-1)/2 = 0.5$, i.e., it is equivalent to the graph of a random walk. We will investigate later to which extent this result is correct for dislocations.

To investigate the wandering dynamics, we consider an initially straight elastic line moving under the influence of thermal forces in an over-damped manner with a drag coefficient per unit length $B$. For a  nearly straight line of line energy ${\cal T}$, the wandering dynamics under the influence of thermal Langevin forces is described by the annealed Edwards-Wilkinson equation
\begin{equation}
B \frac{d y}{d t} = {\cal T}\frac{\partial^2 y}{\partial x^2} + f_T (x,t)
\label{eq:aedwilk}
\end{equation}
where $f_T$ is a Gaussian random force with the correlation function
\begin{equation}
\langle f_T(x,t) f_T(x',t') \rangle = 2 k_{\rm B} T B \delta(x-x') \delta (t-t').
\end{equation}
The linearity of the governing equation \eqref{eq:aedwilk} allows for analytical solution in Fourier space, see e.g. \cite{Nattermann1992_PRA}. The evolution equation of the Fourier modes has the structure of an Ornstein-Uhlenbeck process; its solutions are Gaussian random variables with the correlation function
\begin{equation}
\langle A(q,t) A(q',t') \rangle = \frac{k_B T}{ {\cal T}} \frac{1}{q^2} \left(1 - \exp\left[ - \frac{2 {\cal T}}{B} q^2 t\right ] \right) \delta(q - q')
\end{equation}
or in terms of the wave number
\begin{equation}
\langle A(n,t) A(n',t') \rangle = \frac{2 k_B T}{ {\cal T} L} \left(\frac{L}{\pi n}\right)^2 \left(1 - 
\exp\left[ - \frac{2 \pi^2 {\cal T}}{BL^2} n^2 t\right ] \right) \delta_{nn'}
\end{equation}

\subsection{Spatially fluctuating forces}

If the elastic line is subject to spatially fluctuating but temporally fixed forces, the evolution is given by the quenched Edwards-Wilkinson equation 
\begin{equation}
B \frac{d y}{d t} = {\cal T}\frac{\partial^2 y}{\partial x^2} + f_{\rm q} (x,y) + f_{\rm ext}
\label{eq:qedwilk}
\end{equation}
where $f_{\rm ext}$ is an externally applied driving force per unit length and the random force $f_{\rm q}$ has the correlation function
\begin{equation}
\langle f_{\rm q}(x,y) f_{\rm q}(x',y') \rangle = \hat{f}^2 \xi \delta(x-x') \phi(y-y').
\end{equation}
where $\hat{f}$ is a characteristic force per unit length created by a random potential of correlation length $\xi$ and it is understood that we consider the line on scales well above the correlation length. The function $\phi(y)$ where $\phi(0) = 1, \int \phi dy = \xi$ describes short-range correlations of the random force in the $y$ direction of motion of the line. Owing to the $y$ dependence of the random force, \eqref{eq:qedwilk} is intrinsically nonlinear. 

To understand the energy scales associated with \eqref{eq:qedwilk}, let us consider a bulge of length $L$ and width $\xi$ on 
an otherwise straight dislocation. The characteristic restoring force due to line tension is then easily estimated as $F_{\rm LT} \approx {\cal T} \xi/L$. The total random force acting on the bulge is a random variable with zero mean and standard deviation  $F_{\rm R} = \hat{f} \sqrt{\xi L}$. Comparing the random and the restoring force defines a characteristic length, the so-called pinning length, given by 
\begin{equation}
L_{\rm p} = \left(\frac{\cal T}{\hat{f}}\right)^{2/3} \xi^{1/3}
\label{eq:lpin}
\end{equation}
Below this length, the restoring force prevails, i.e., the line is bound to remain essentially straight. On scales above this length, the line will in the absence of an external driving force wander by an amount of the order of $\xi$ to reach the nearest 
energy minimum. For an initially straight line parallel to the $y$ axis this leads to a characteristic line shape where the line 
is straight but inclined on scales below $L_{\rm p}$, and exhibits irregular fluctuations with a standard deviation of the order of $\xi$ while keeping its original orientation on scales above $L_{\rm p}$. This allows us to obtain estimates of the pinning length directly from the shape of the relaxed line.

Averaging the pinning force over the pinning length defines a characteristic force per unit length. If the external force is smaller than this effective pinning force, the line is likely to be trapped into some metastable configuration. If the external force is large enough to make all metastable configurations disappear, the line will move indefinitely (depinning). The critical force for depinning can be estimated by comparing the external force to the pinning force, averaged over the pinning length. This gives
\begin{equation}
f_{\rm ext,c}= \hat{f} \sqrt{\frac{\xi}{L_{\rm p}}} = \hat{f}^{4/3} {\cal T}^{-1/3} \xi^{1/3}.
\label{eq:fpin}
\end{equation}
In the context of dislocation theory, this argument corresponds to the weak pinning limit. An interesting aspect is that, if the pinning length is known, we may relate the critical force directly to the pinning length by eliminating the characteristic force $\hat{f}$ from Eqs. (\ref{eq:lpin}) and (\ref{eq:fpin}) to obtain
\begin{equation}
f_{\rm ext,c}= \frac{{\cal T}\xi^{1/3}}{L_{\rm p}^2}.
\label{eq:fpinL}
\end{equation}
As we shall demonstrate, this relationship allows to obtain, in the weak pinning regime, rough estimates of the pinning force in a numerically very efficient manner. 

\section{Discrete Stochastic Dislocation Dynamics}
\label{sec:3}

\subsection{Description of model}
\label{sec:3.1}
In the present model which focuses on core effects, a non singular theory \cite{Cai2006_JMPS,Arsenlis2007_MSMSE} is considered. After discretization of the dislocation into nodes connected by segments, the equation of motion of dislocation node $i$ is escribed by the following equation:
\begin{equation}\label{equ:1}
m_{0}L_{i} \bm{a}_{i}=-BL_{i} \bm{v}_{i} + \bm{F}_{i}^{\rm disloc} + \bm{F}_{i,T} + \bm{F}_{i}^{\rm a} + \bm{F}_{i}^{\rm q}.
\end{equation}
Here $m_{0}\approx\rho b^{2}/2$ is the effective dislocation mass per unit length, , $\rho$ and $b$ are the material mass density and magnitude of Burgers vector, respectively; $L_{i}$ is an effective length associated with node $i$ due to dislocation segments connecting to it; $B$ is the drag coefficient per unit length; $ \bm{f}_{i}^{\rm disloc} $ includes the self force $\bm{f}_{i}^{\rm self}$, core force $\bm{f}_{i}^{\rm core} $ and the force from dislocation segment-segment interactions; $\bm{f}_{i,T}$ is a Langevin force which accounts for thermal fluctuations; $\bm{f}_{i}^{\rm a}$ is the force due to the externally applied stress and $\bm{f}_{i}^{\rm q}$ is the force due to a spatial force field that may be used to model the influence of atomic disorder. 

The Langevin force is a stochastic force of magnitude (see \ref{app:1})
\begin{equation}
\bm{F}_{i,T} =\Lambda_{i,j}\sqrt{\frac{2Bk_{B}TL_{i}}{\Delta t_{j}}}\bm{s}_{i},
\end{equation}
where $\Lambda_{i,j}$ is a normal distributed Gaussian random number; $k_{B}$ is Boltzmann's constant; $T$ is absolute temperature; $\Delta t_{j}$ is the increment of time at simulation step $j$, and the unit vector $\bm{s}_{i}$ denotes the effective glide direction of node $i$. 

\begin{figure*}[t]
	\centering
	\subfigure[]{\includegraphics[width=3.3in]{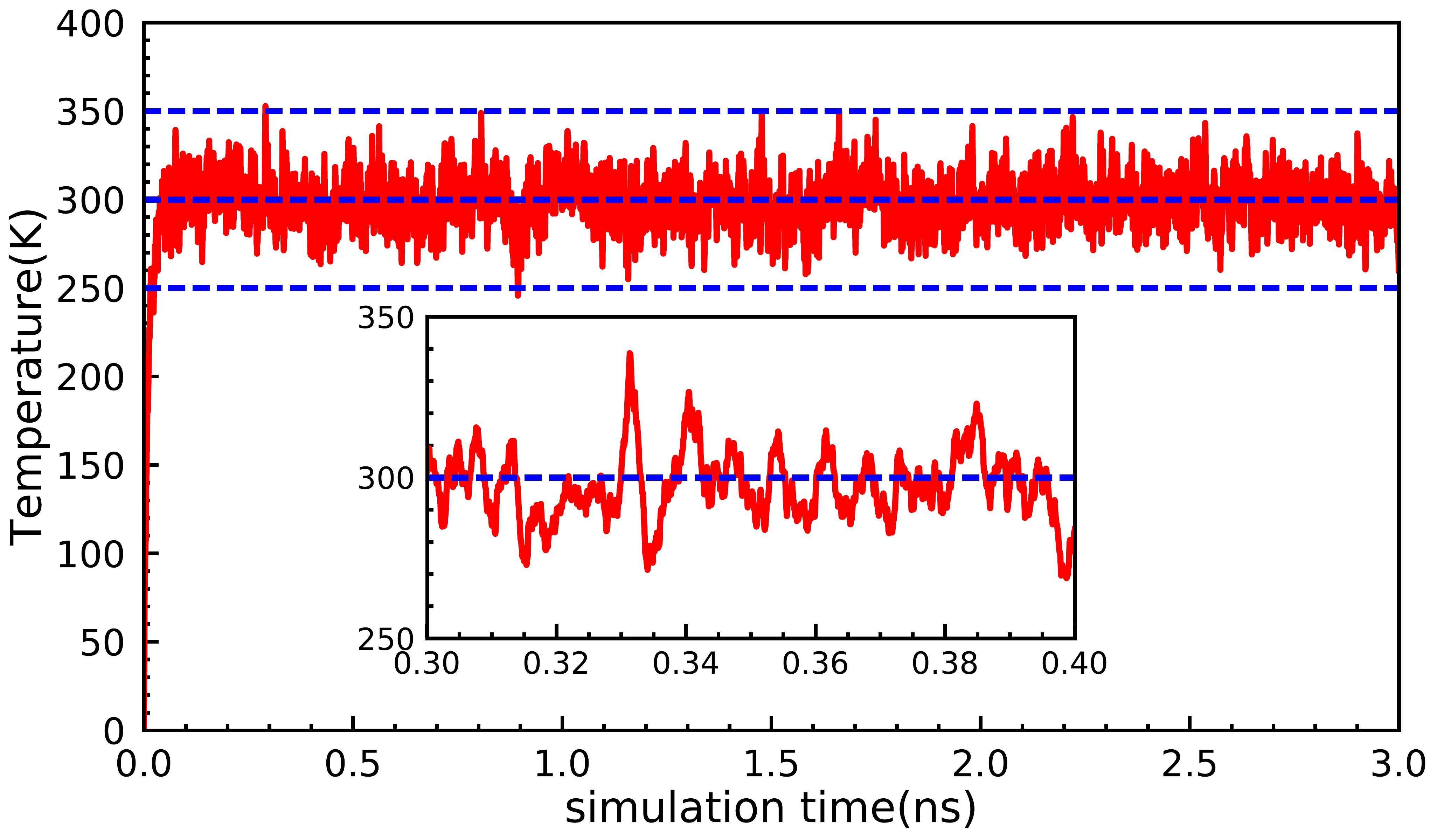}}
	\subfigure[]{\includegraphics[width=3.3in]{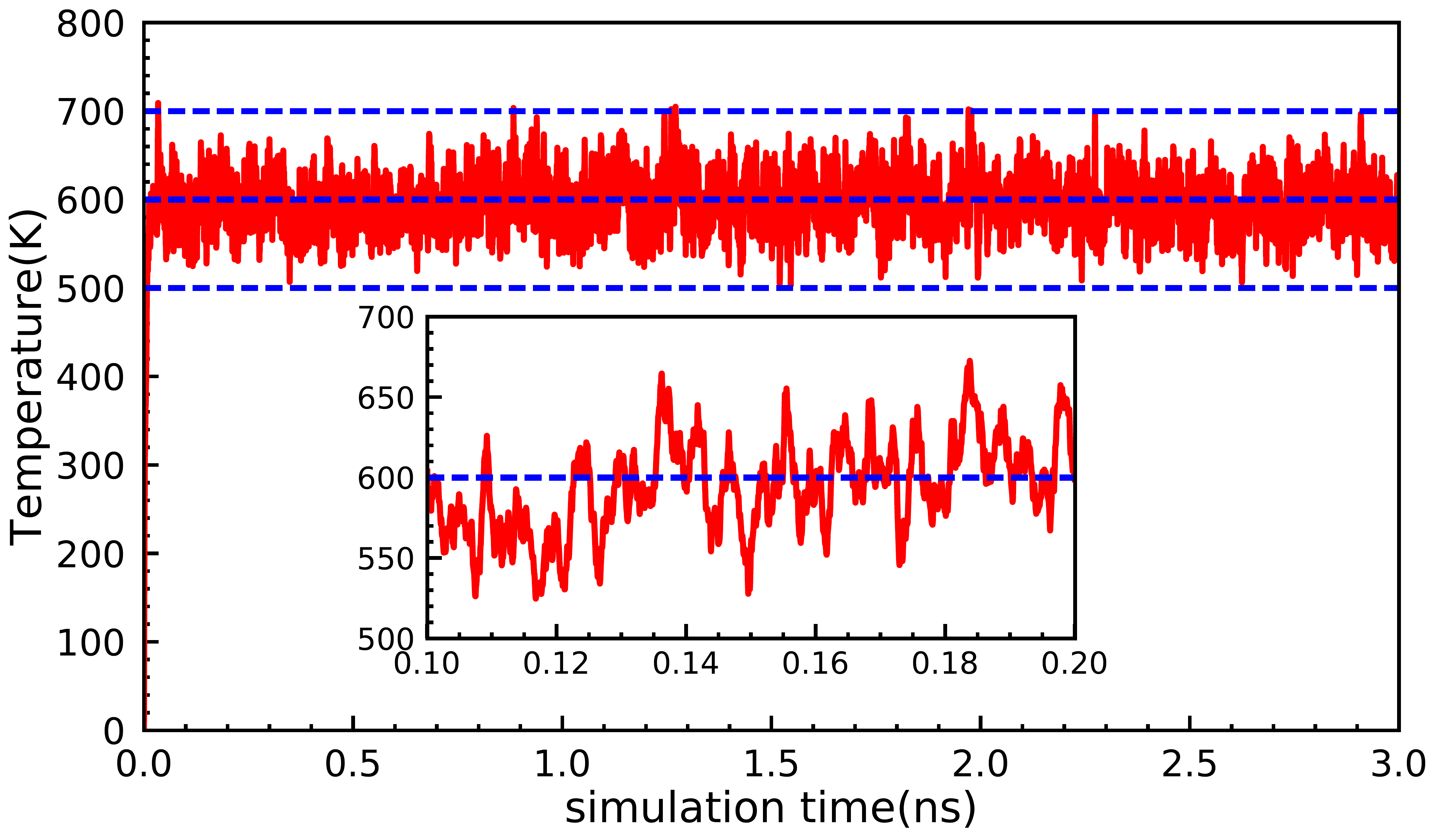}}	
	\subfigure[]{\includegraphics[width=3.3in]{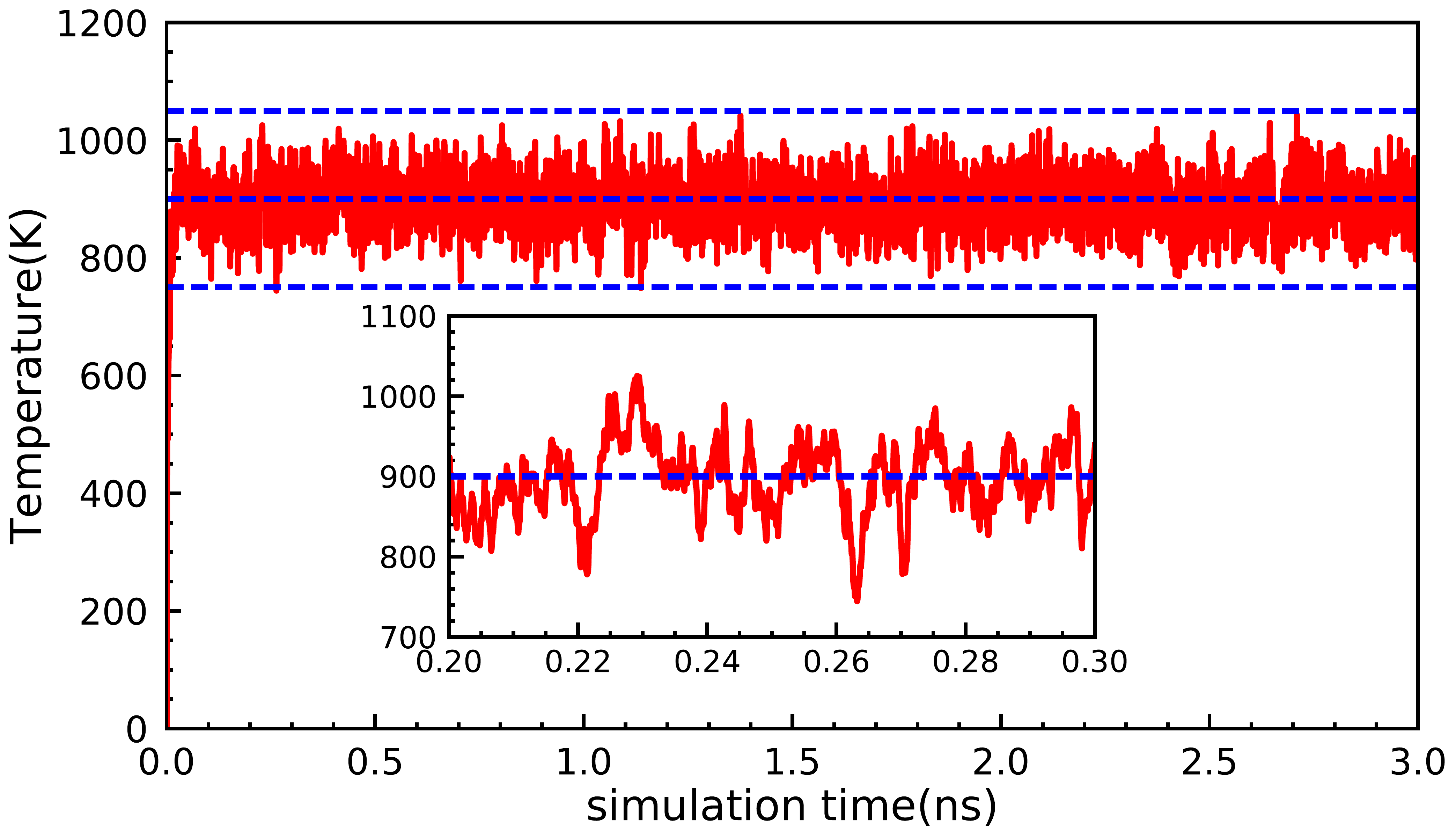}}
	\subfigure[]{\includegraphics[width=3.3in]{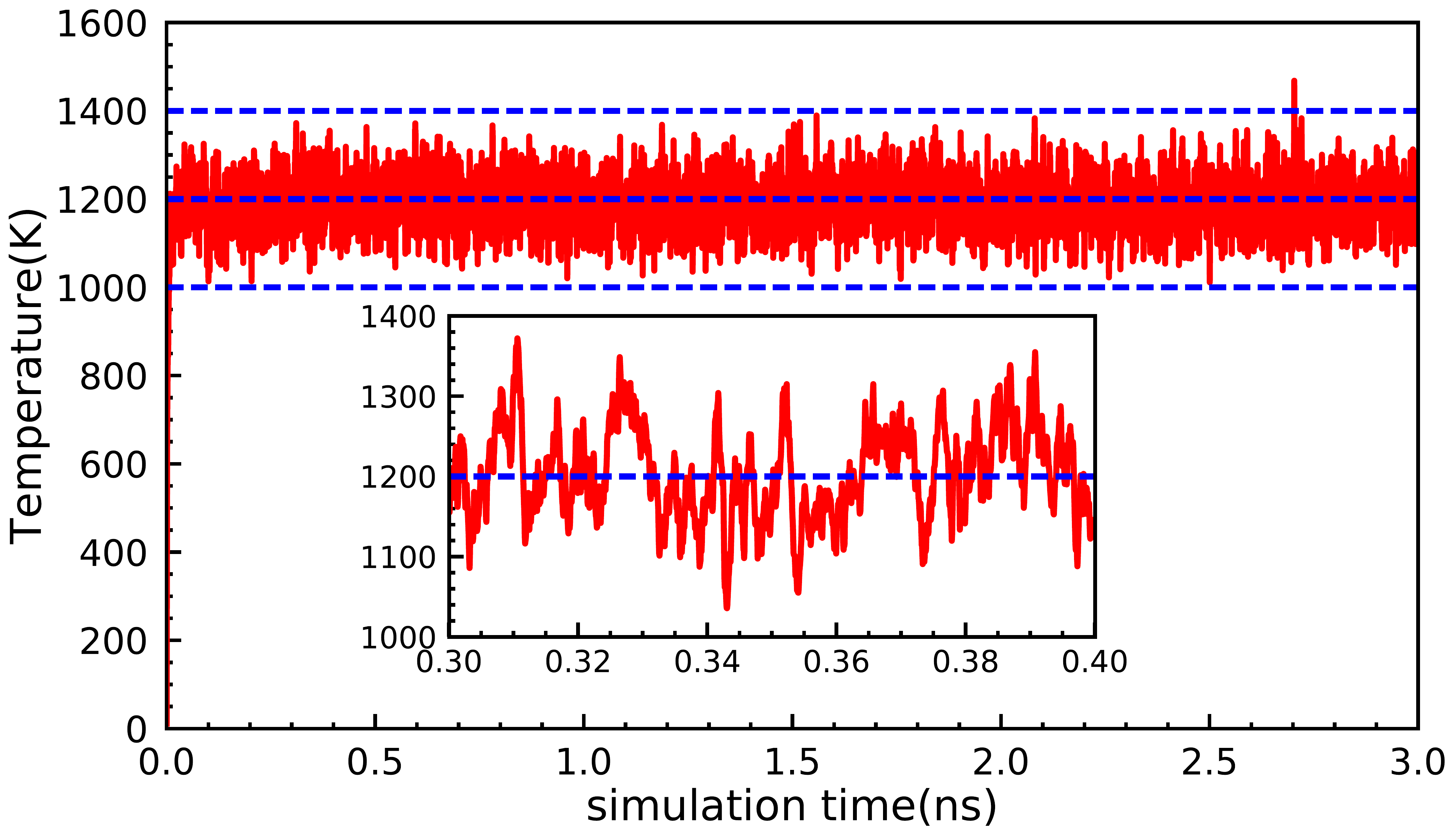}}
	\caption{Kinetic temperature as function of time: (a) 300K, (b) 600K, (c) 900K, (d) 1200K.}
	\label{fig:kinetic_temperature}
\end{figure*}

\begin{table}
	\caption{Temperature dependence of drag coefficient, after \cite{Hiratani2002_JEMT}.}
	\centering
	\begin{tabular}{|c|c|}
		\hline Temperature(K) & Drag coefficient($\mu$Pa$\bm{\cdot}$s) \\
		\hline  300				&  30			\\
		\hline  600				&  60			\\
		\hline  900				&  90			\\
		\hline 	1200			&  120			\\
		\hline
	\end{tabular}
	\label{table:1}
\end{table}

\begin{figure*}[t]
	\centering
	\subfigure[]{\includegraphics[width=3.3in]{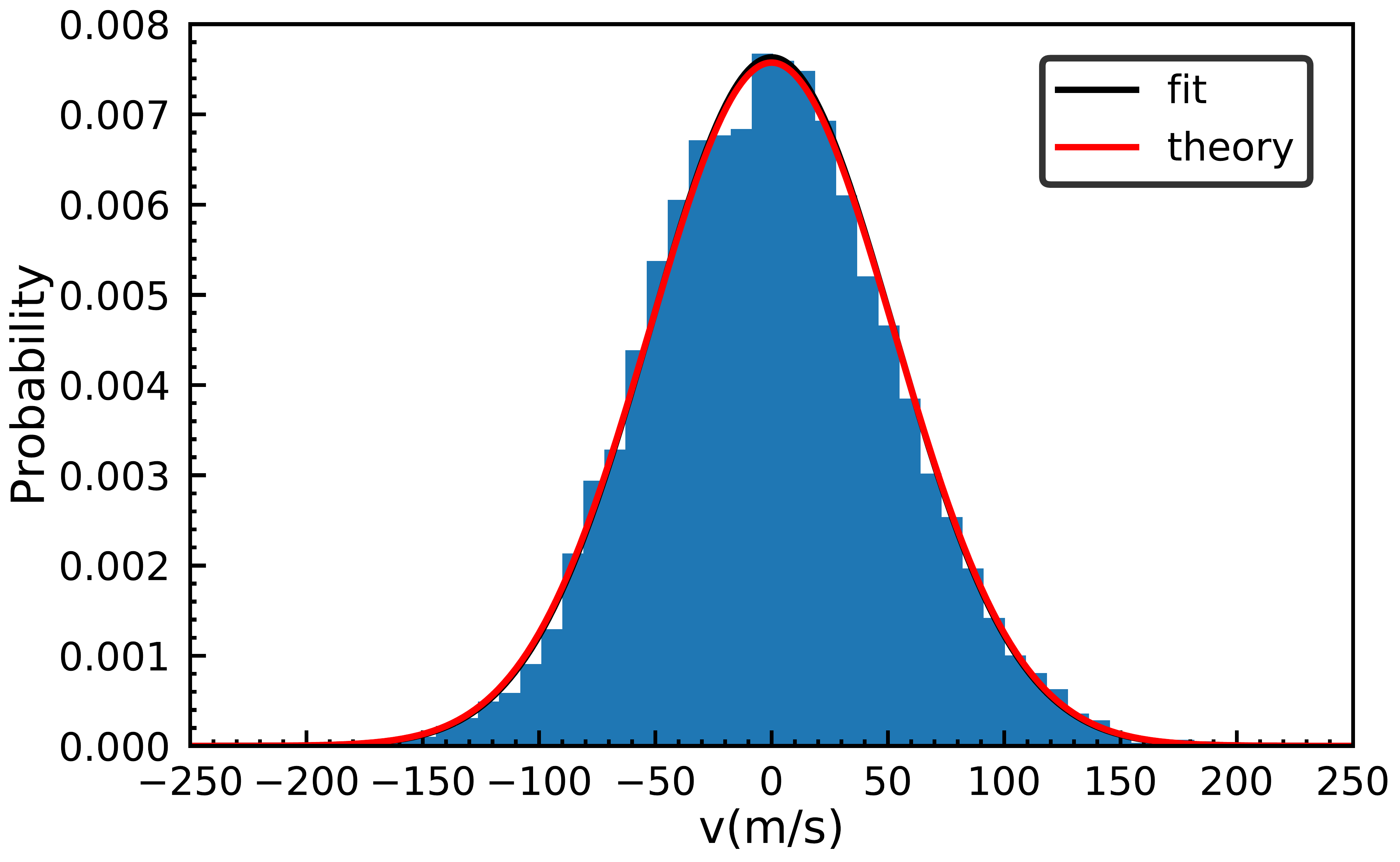}}
	\subfigure[]{\includegraphics[width=3.3in]{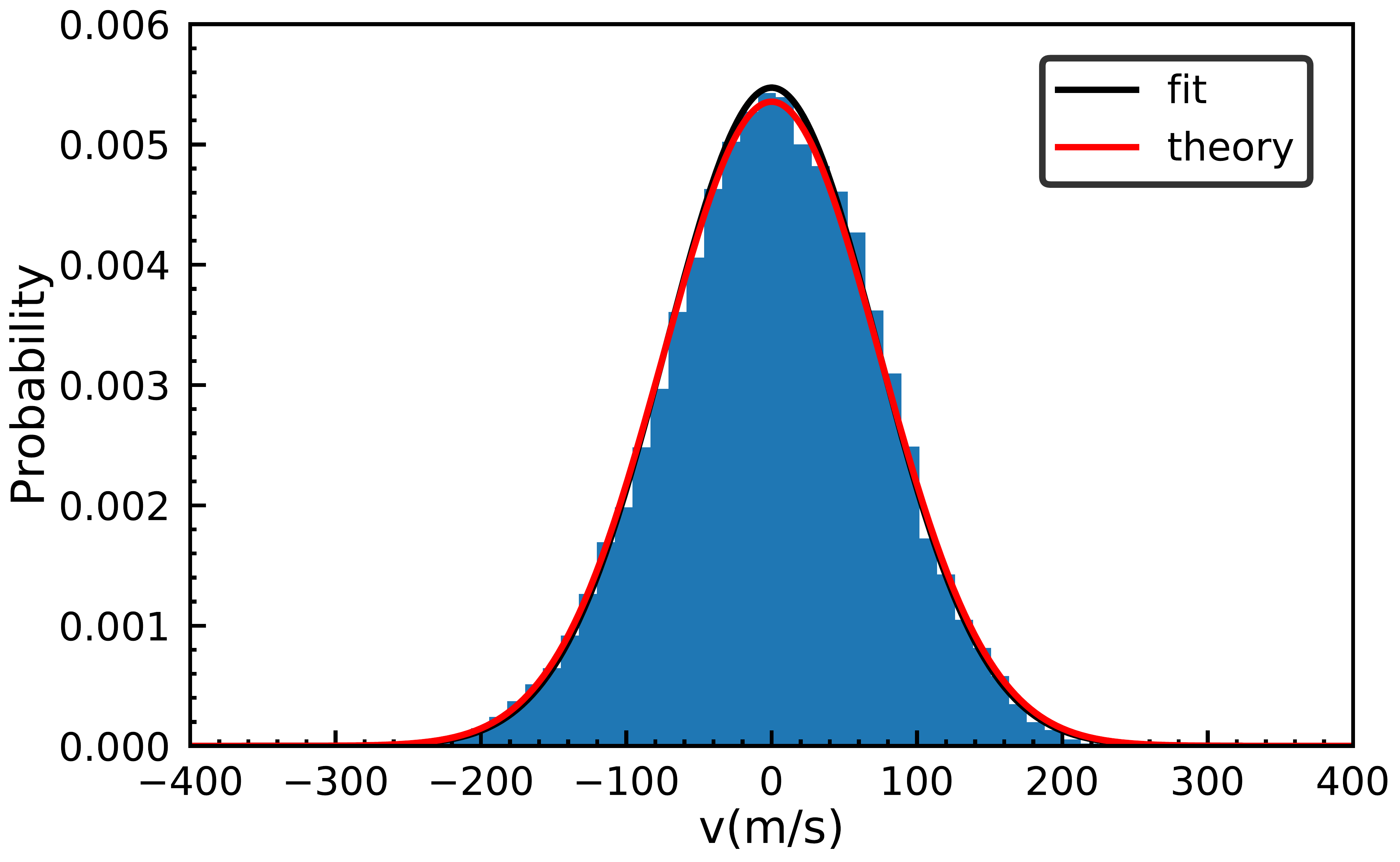}}
	\subfigure[]{\includegraphics[width=3.3in]{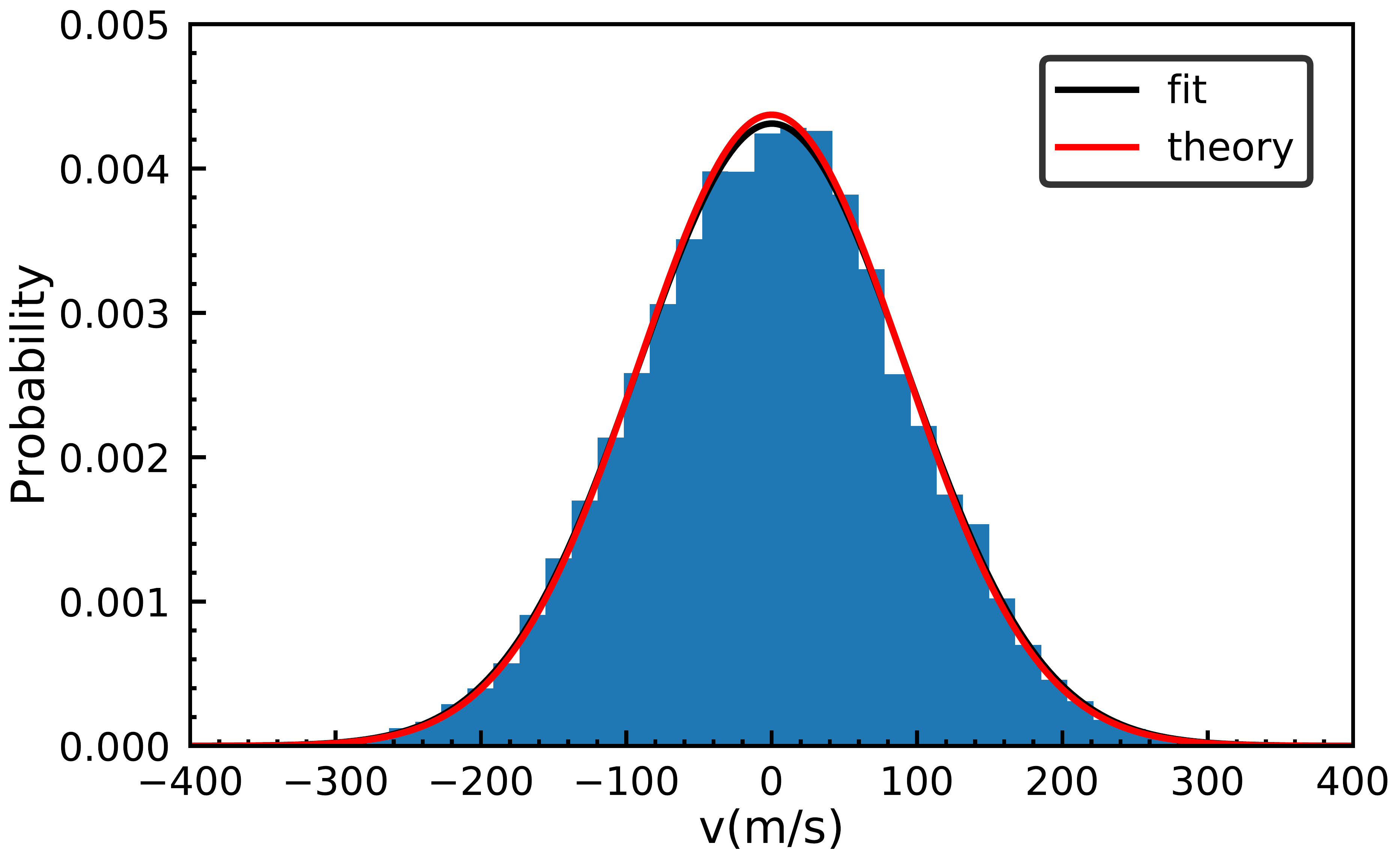}}
	\subfigure[]{\includegraphics[width=3.3in]{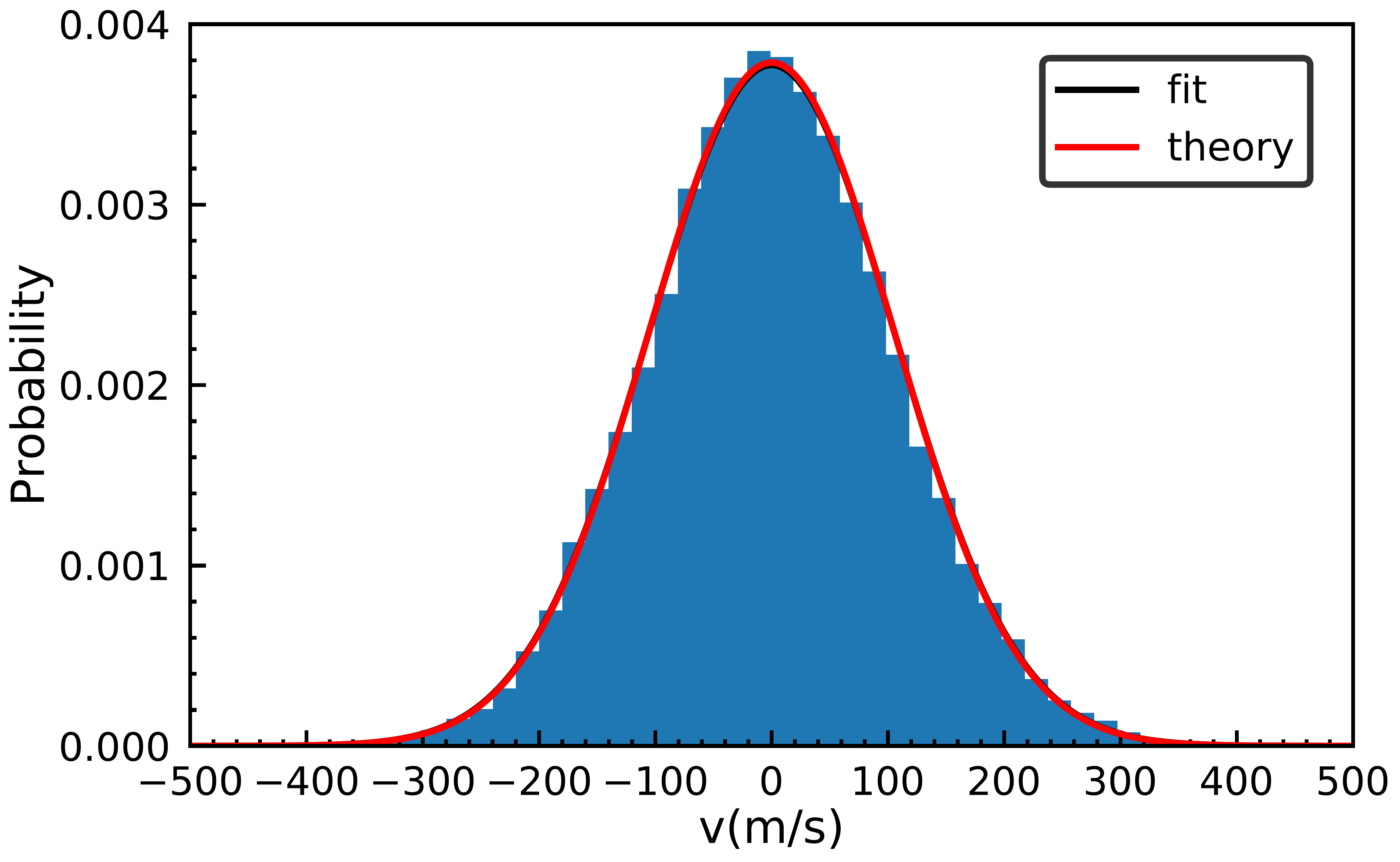}}
	\caption{Distribution of velocity: (a) 300K, (b) 600K, (c) 900K, (d) 1200K.}
	\label{fig:velocity_distribution}
\end{figure*}

\begin{table*}[t]
	\caption{Simulated thermal quantities.}
	\centering
	\begin{tabular}{|c|c|c|c|c|>{\centering}p{1.2cm}|c|c|c|}
		\hline \multirow{2}{*}{given temperature(K)} & \multicolumn{2}{c|}{temperature(K)}  & \multicolumn{2}{c|}{TSD(m/s)} & \multicolumn{2}{c|}{simulated SD(m/s)} & \multicolumn{2}{c|}{$t_{equ}$(ps)}\\
		\cline{2-9}
		& (i) &(ii) &(i) &(ii) &(i) &(ii) &(i) &(ii)\\
		\hline 300 & 299 & 299 &52.7 &16.6 &52.3 &17.2 & $\sim$31.6&$\sim$36.5 \\
		\hline 600 & 596 & 596 &74.5 &23.5 &72.9 &23.6 & $\sim$21.2& $\sim$16.7\\
		\hline 900 & 894 & 896 &91.2 &28.8 &92.5 &29.8 & $\sim$13.1& $\sim$15.9\\
		\hline 1200 & 1189& 1189&105.3 &33.3 &105.7 & 32.4& $\sim$9.0& $\sim$8.2\\
		\hline
	\end{tabular}
	\label{table:2}
\end{table*}

\subsection{Validation: thermal equilibrium properties}
\label{sec:3.2}
To validate our model implementation, we consider an initially straight edge dislocation which we assume to run in $x$ direction and move in the $xy$ plane under the influence of thermal forces. After sufficient simulation time, once thermal equilibrium has been reached, the mean kinetic energy for each degree of freedom should be $k_{B}T/2$ and the distribution of nodal velocities along the local directions $\bm{s}_{i}$ should follow Maxwell's distribution with a theoretical standard deviation (SD) of $\sqrt{k_{B}T/(m_{0}L_{i})}$. Two different simulations are considered in this part: (i) an edge dislocation line with fixed endpoints which is discretized into 1000 dislocation segments with segment length 20$b$; (ii) a similar edge dislocation line with non-equidistant nodes and $L_{i}$ changing from 20$b$ to 200$b$. The time step $\Delta t_{j}$ is chosen constant and equals 50fs. For larger time steps, the mean kinetic energy and velocity distribution can be used to determine whether the selection is suitable or not. A series of temperatures (300K, 600K, 900K and 1200K) are simulated to check the correctness of the Langevin force implementation. Material parameters are taken from \cite{Hiratani2002_JEMT} in order to compare with these simulations, in which Langevin force were applied on dislocation segments rather than nodes. The shear modulus and Poisson ratio are $\mu$=54.6 GPa and $\nu$=0.324, respectively; the magnitude of Burgers vector $b$=0.256nm and mass density is $\rho$=8900kg/m$^{3}$. The drag coefficient is assumed proportional to temperature and listed in Table \ref{table:1}.

Kinetic temperatures and velocity distributions for different Langevin temperatures  are shown in Figs. \ref{fig:kinetic_temperature} and \ref{fig:velocity_distribution}, respectively for a dislocation with equi-distant nodes. The principle of equipartition is observed in Fig. \ref{fig:kinetic_temperature} and the distribution of velocity about a specified node follows a Maxwell distribution with simulated SD close to the theoretical value $\sqrt{k_{B}T/(m_{0}L_{i})}$ as shown in Fig. \ref{fig:velocity_distribution}. Simulated temperature, theoretical value of standard deviation (TSD), simulated SD and time $t_{\rm equ}$ for thermal equilibration are shown in Table \ref{table:2} (for dislocations with equi-distant nodes (case (i)) and for dislocations with variable node spacing (case (ii)), in the latter case, the SD and TSD data refer to the node with the largest associated effective length. As we can see, simulated temperatures are very close to the given temperatures (maximum relative error is within 1\%) and the simulated SD of velocity distribution has matches the theoretical values. A non-equidistant node distribution scheme has no detrimental influence on the agreement between theory and simulation. $t_{\rm equ}$ is mainly influenced by drag coefficient and decreases with increasing drag coefficient. The influence of dislocation segment length is also studied. Simulated temperature, TSD, simulated SD and $t_{\rm equ}$ for different dislocation segment lengths in a simulation of type (i) are shown in Table \ref{table:3} where $\Delta t$=50fs and $T$=1200K. The table shows that different dislocation segment lengths have no influence on simulated temperature and thermal equilibration time decreases with increased dislocation length, which is the opposite of the result obtained when Langevin forces are applied on dislocation segments  \cite{Hiratani2002_JEMT}.
\begin{table*}[b]
	\centering
	\caption{Influence of different dislocation segment lengths.}
	\begin{tabular}{|c|c|c|c|c|}
		\hline Length($b$) & simulated temperature(K) & TSD(m/s) & simulated SD(m/s) &$t_{equ}$(ps)\\
		\hline 100		   &    1188      & 47.1    & 47.0  & $\sim$8.7\\
		\hline 200         &    1190      & 33.3    & 32.3  & $\sim$8.1\\
		\hline 400         &    1190      & 23.5    & 23.7  & $\sim$5.8 \\
		\hline 800         &    1185      & 16.6    & 15.4  & $\sim$4.4 \\
		\hline
	\end{tabular}
	\label{table:3}
\end{table*}

\section{Simulation results}
\label{sec:4}

\subsection{Spatio-temporal roughening of a dislocation line under influence of thermal forces}
\label{sec:4.1}

We simulate an edge dislocation of length $2 \times 10^4 b$ pinned at its endpoints. The line direction is identified with the $x$ direction of a Cartesian coordinate system, the glide plane is the $xy$ plane. The system is assumed to be of infinite extension in $z$ direction and periodic boundary conditions are imposed on a simulation cell of extension $2 \times 10^4 b$ in $x$ and $2 \times 10^4 b$ in $y$ direction. The dislocation is discretized into 1000 nodes with an inital spacing of $20b$.

In our investigation of the spatio-temporal behavior of the dislocation under the influence of thermal forces, we first consider a dislocation that has been evolving for sufficient time to establish thermal equilibrium. In this case the amplitudes of its Fourier modes fulfill Eq. \eqref{eq:intens}. 
\begin{figure}[htb]
	\centering
		\includegraphics[width=3.3in]{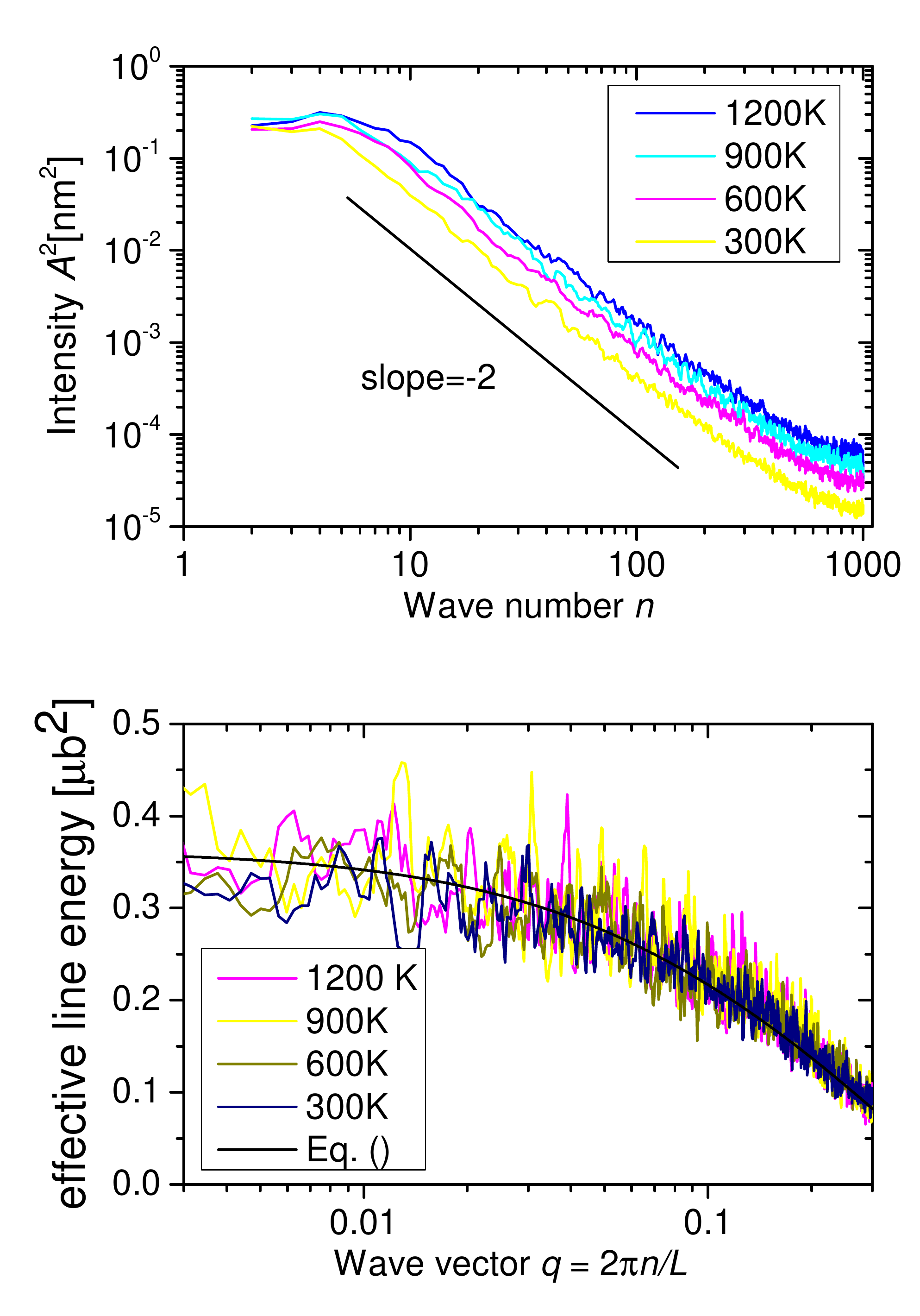}
		\caption{Top: Fourier mode amplitudes of a thermally equilibriated dislocation against wave number; dashed line: $A_n^2 \propto n^{-2}$ as expected for an elastic line according to \eqref{eq:intens}; bottom: effective line energy as determined
		from \eqref{eq:linetens}, full: guide to the eye.\label{fig:eqdis}}
\end{figure}
The amplitudes of the Fourier modes of the equilibrated dislocation are shown in Figure \ref{fig:eqdis}, top. We observe that, for short wavelengths (large wave numbers), characteristic deviations occur from the $A_n^2 \propto n^{-2}$ behavior expected according to Eq. \ref{fig:eqdis}. Of course, in our simulations the self-interaction of a dislocation line is described by a non-local interaction kernel which is much more complex than a simple elastic line. So the question arises whether we can understand the deviations and quantify them. 

To this end, we first use \eqref{eq:linetens} to determine a wave number dependent line energy. This is shown in Figure \ref{fig:eqdis}, bottom. We can see that, for long-wavelength fluctuations, the data can be well described by an approximately constant line energy. However, on short scales (large wave numbers) the effective line tension seems to be decreasing. To understand this behavior, we have, in Appendix A, calculated the Fourier transform of the non-singular dislocation self interaction kernel \cite{Cai2006_JMPS,Arsenlis2007_MSMSE}, both for an isolated dislocation and for a periodic array of rigidly coupled dislocations as implied in the present simulations by our boundary conditions.

\begin{figure}[htb]
	\centering
		\includegraphics[width=3.3in]{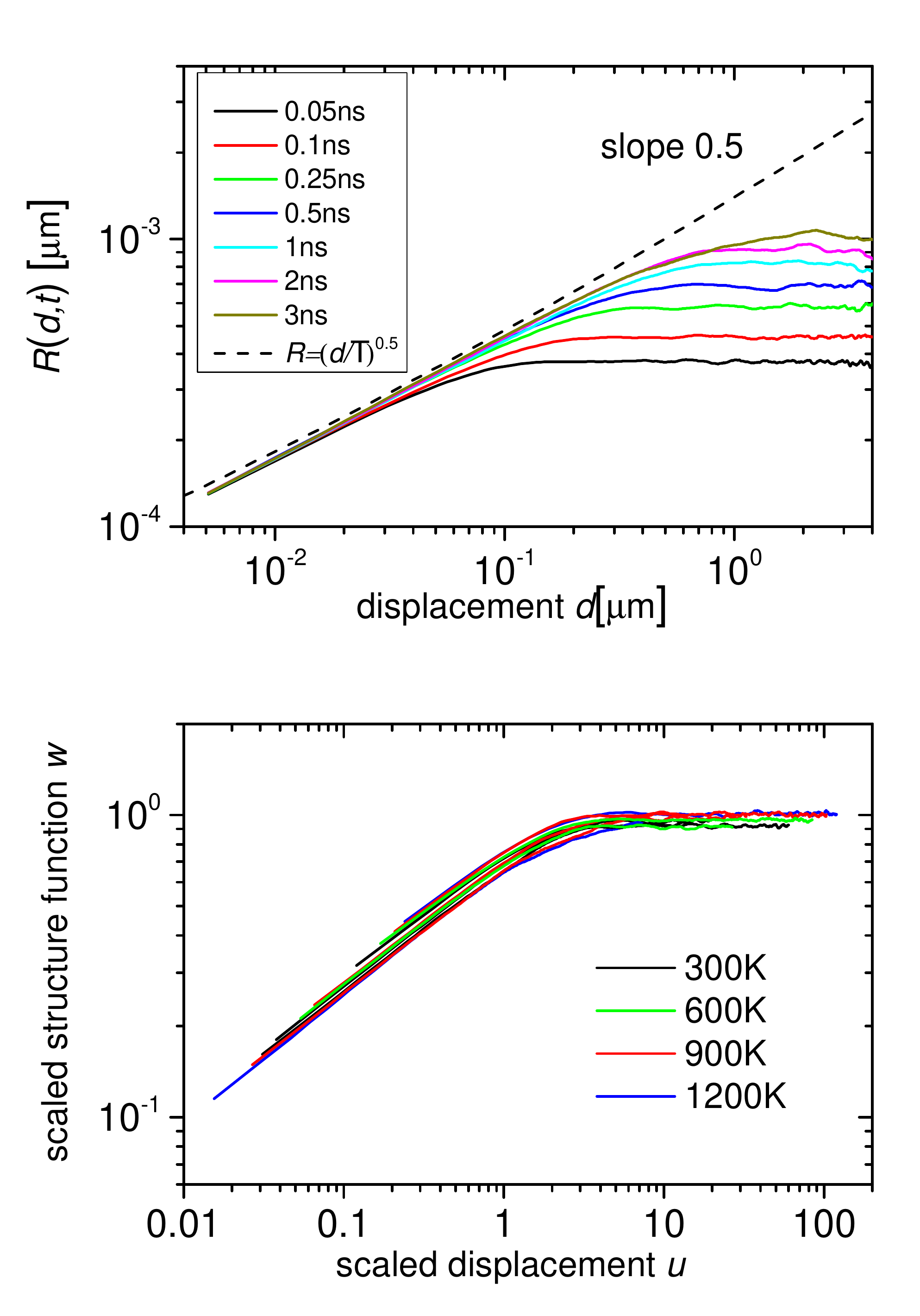}
	\caption{\label{fig:rougheningT}Top: time evolution of the structure function $R(d,t)$, simulation at $T$=300K; dashed line: $R \propto
	(d/{\cal T}(d)^{0.5}$, average over 10 simulations; bottom: scaled structure function $w = R (k_{\rm B}T)^{-0.5} B^{0.25} {\cal T}^{0.25} t^{-0.25}$ as function of scaled displacement $u = d({\cal T} t/B)^{-0.5}$; data for different temperatures as shown in legend, data at each temperature for times $t=0.05ns, t= 0.5ns, t=3ns$.}
\end{figure}
In a next step, we investigate the transient dynamics before the dislocation has reached thermal equilibrium, starting from an initially straight dislocation line configuration which, under the influence of thermal fluctuations, gradually develops a self affine shape $y(x)$. We statistically characterize this process by the structure factor $R(d,t)$ defined as:
\begin{equation}
\label{equ:3}
R(d,t)=\left\langle \left\langle \left| y(x,t)-y(x+d,t)\right|\right\rangle_{x}\right\rangle _{N}
\end{equation}
where $\left\langle \right\rangle_N $ means average over an ensemble of $N$ simulations and $\left\langle \right\rangle_x$ the average over all $x$ positions for which $y(x+d,t)$ can be computed. The time evolution of $R(d,t)$ is shown in Figure \ref{fig:rougheningT} for $T=300$K. For a given time, the function $R(d,t)$ exhibits two regimes: At small $d$, thermal equilibrium has been reached, and the line has a self affine shape where the structure function is time independent and approximately follows the theoretically expected scaling for an elastic line, $R \propto (d k_{\rm B} T/{\cal T})^{\zeta}$ with $\zeta = 0.5$ \cite{Nattermann1992_PRA}. The apparent deviations from slope 0.5 on the double-logarithmic plot can be accounted for by noting that the effective line energy of the dislocation decreases on small scales. Setting ${\cal T} \propto \ln(d/b)$ produces a good representation of the data. 

\begin{figure}[hbt]
	\centering
		\includegraphics[width=3.3in]{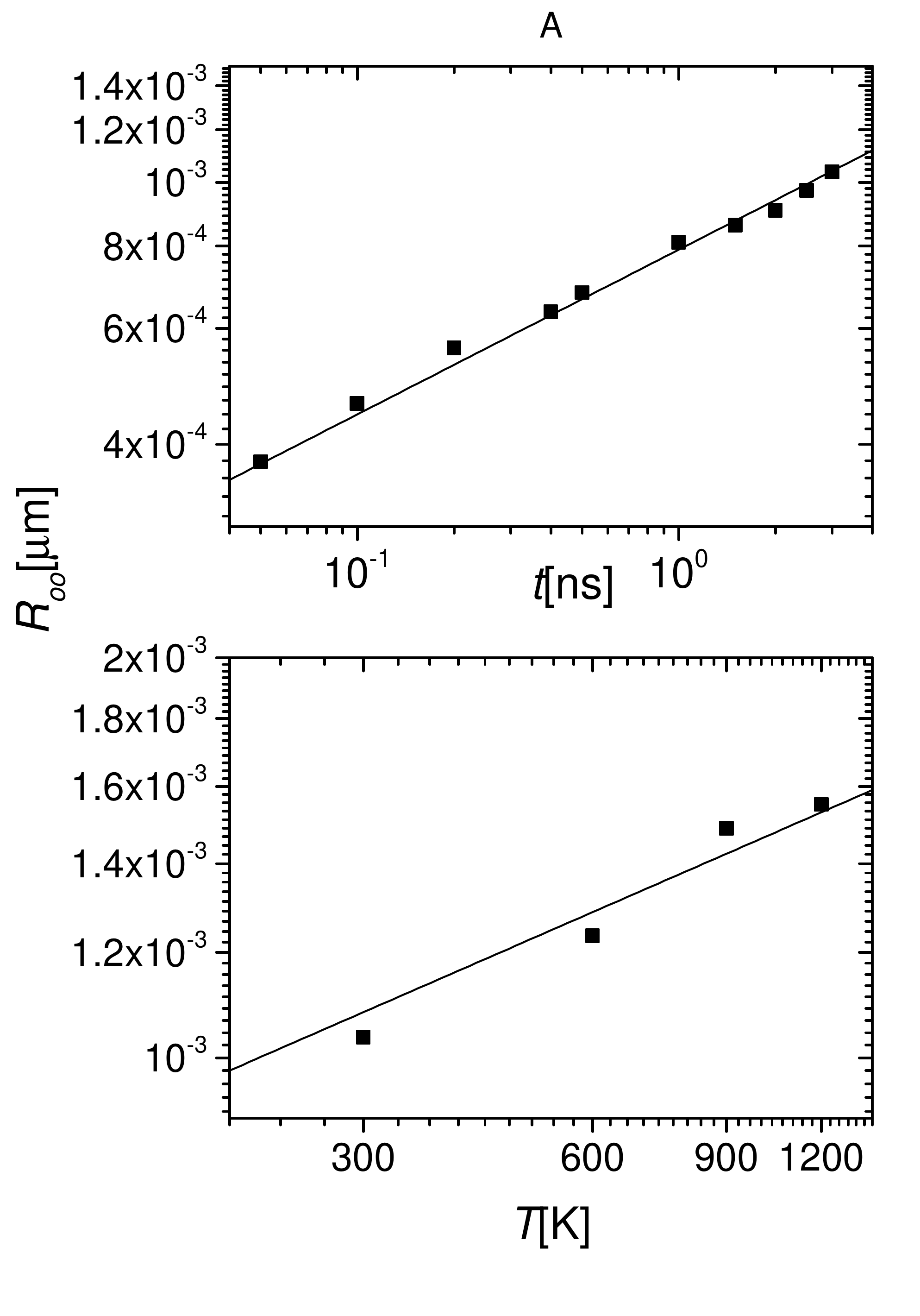}
	  \caption{Top: time dependence of the asymptotic structure factor $R_{\infty}$ at a temperature of 1200 K, full line: slope 0.25
		of the double-logarithmic plot; bottom: temperature dependence of $R_{\infty}$ at time $3ns$, full line: slope 0.25
		of the double-logarithmic plot.	\label{fig:rtimetemp}}
\end{figure}

The overall behavior of $R(d,t)$ can be described by the equation
\begin{equation}
R(d,t) = \left(\frac{d k_{\rm B} T}{{\cal T}}\right)^{\zeta} \Phi\left(\frac{\cal T}{B} \frac{t}{d^z}\right)
\end{equation}
with the roughness exponent $\zeta = 0.5$ and the dynamic exponent $z=2$ for the elastic line under thermal forces. The scaling
function $\Phi$ has the following properties: $\Phi(x) \to \Phi_{\infty}\; {\rm for}\; x \to \infty, \Phi(x) \propto x^{\zeta/z}\;
{\rm for}\; x \to 0$. This implies that, for sufficiently large $d$, the structure function approaches an approximately $d$ independent value $R_{\infty} \propto T^{\zeta} B^{-\zeta/z} {\cal T}^{\zeta(1-z)/z} t^{\zeta/z}$. With $\zeta = 0.5$, $z = 2$, and a linear temperature dependency $B \propto T$, we find that $R_{\infty} \propto T {0.25} t^{0.25}$. Figure \ref{fig:rtimetemp} shows that the time- and temperature dependence of $R_{\infty}$ is in good agreement with this expectation. 

A generic representation of the roughening behavior at all times and all temperatures is obtained by introducing the new variables $u = d({\cal T} t/B)^{-1/z}$ and $w = R (k_{\rm B}T)^{-\zeta} B^{\zeta/z} {\cal T}^{\zeta(z-1)/z} t^{-\zeta/z}$ in which the structure factor has the universal form
\begin{equation}
w = u^{\zeta} \Phi(u^{-z}).
\end{equation} 
This is illustrated in Figure \ref{fig:rougheningT}, bottom, which shows a compilation of structure factor curves pertaining to different temperatures and roughening times after re-scaling to the universal variables $u$ and $w$. The plot demonstrates the existence of an underlying generic scaling curve that represents the evolving shapes at all times and temperatures.

\subsection{Roughening and pinning in a spatially random force field}
\label{sec:4.2}
In this section, we investigate the interaction of a dislocation with a spatial time-independent force field. Spatially random force fields can be used to model structural disorder on the atomic scale as prominent in random solid solutions and high-entropy alloys. Such disorder is, at low to intermediate temperatures, time independent ('quenched disorder') and creates spatially fluctuating internal stresses and ensuing forces on the dislocations which depend on position and have zero average value. We use a simple model where a random force at the point $(x,y)$ is created by the superposition of forces from randomly located pinning points, each of which creates a Gaussian pinning potential:
\begin{equation}
	f(x,y)=\hat{f}\sum_{i=1}^{N}\frac{(x-x_{i})}{\xi}e^{-\frac{(x-x_{i})^{2}+(y-y_{i})^{2}}{2\xi^{2}}}.
\end{equation}
Here $x_{i},y_{i}$ are the coordinates of pinning center $i$; $N$ is total number of randomly distributed pinning centers; $\xi$=$\sqrt{A/N}$ is the range of the pinning potential which also determines the correlation length of the random pinning force, and $\hat{f}$ defines the characteristic magnitude of the spatial force field. 

\begin{figure}[hbt]
	\centering
	\includegraphics[width=3.3in]{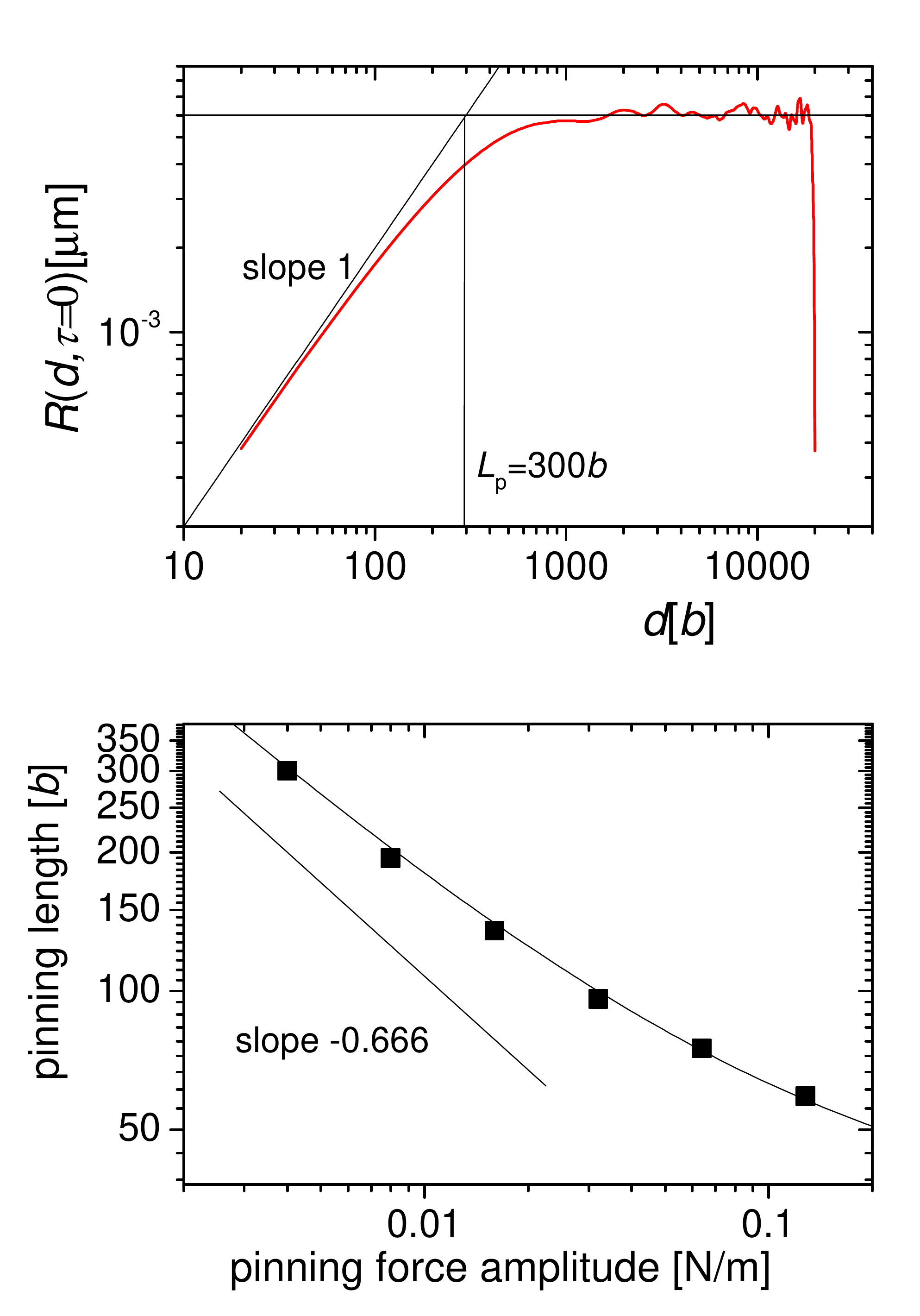}
	\caption{Top: determination of the pinning length, simulation with $\hat{f} = 0.04$N/m at zero applied stress; bottom:
	dependence of pinning length on amplitude $\hat{f}$ of the pinning field.}
	\label{fig:pinlength}
\end{figure}
To quantify the action of the pinning force on the dislocation, we again consider the structure factor $R(d)$ which we evaluate for an ensemble of initially straight dislocations interacting with a pinning potential of range $\xi = 20 b$ and variable strength. During relaxation at zero applied stress, the dislocations develop local roughness as they adjust their line shape to minimize their energy in the random pinning field. The corresponding structure factor is shown in Figure \ref{fig:pinlength}: On small scales one finds an increasing $R(d)$ function with an approximately linear dependency, $R \propto d^{\zeta}$ with $\zeta =1$, as indicative for an inclined dislocation line that has moved under the influence of random forces while remaining locally straight. On large scales the function $R(d)$ is constant as indicative of uncorrelated random displacements of the initially straight line in both directions. A crossover length $L_{\rm p}$ can be constructed as shown in Figure \ref{fig:pinlength} by fitting two curves $R \propto d$ and $R = {\rm const.}$ to the two branches and identifying the crossover length ('pinning length') with the intersection point. 

Results are shown in Figure \ref{fig:pinlength}, top which shows the thus determined pinning length as a function of the strength of the pinning field. As expected according to \eqref{eq:lpin}, the pinning length decreases with increasing pinning field. However, this decrease is only in the regime of weak pinning fields quantitatively described by the theoretical slope of -2/3. At large fields, the pinning length saturates because $L_{\rm p}$ cannot become larger than the physical spacing $\xi$ of the pinning centers. This corresponds to a transition from
weak to strong pinning. 

\begin{figure}
\centering
		\includegraphics[width=3.3in]{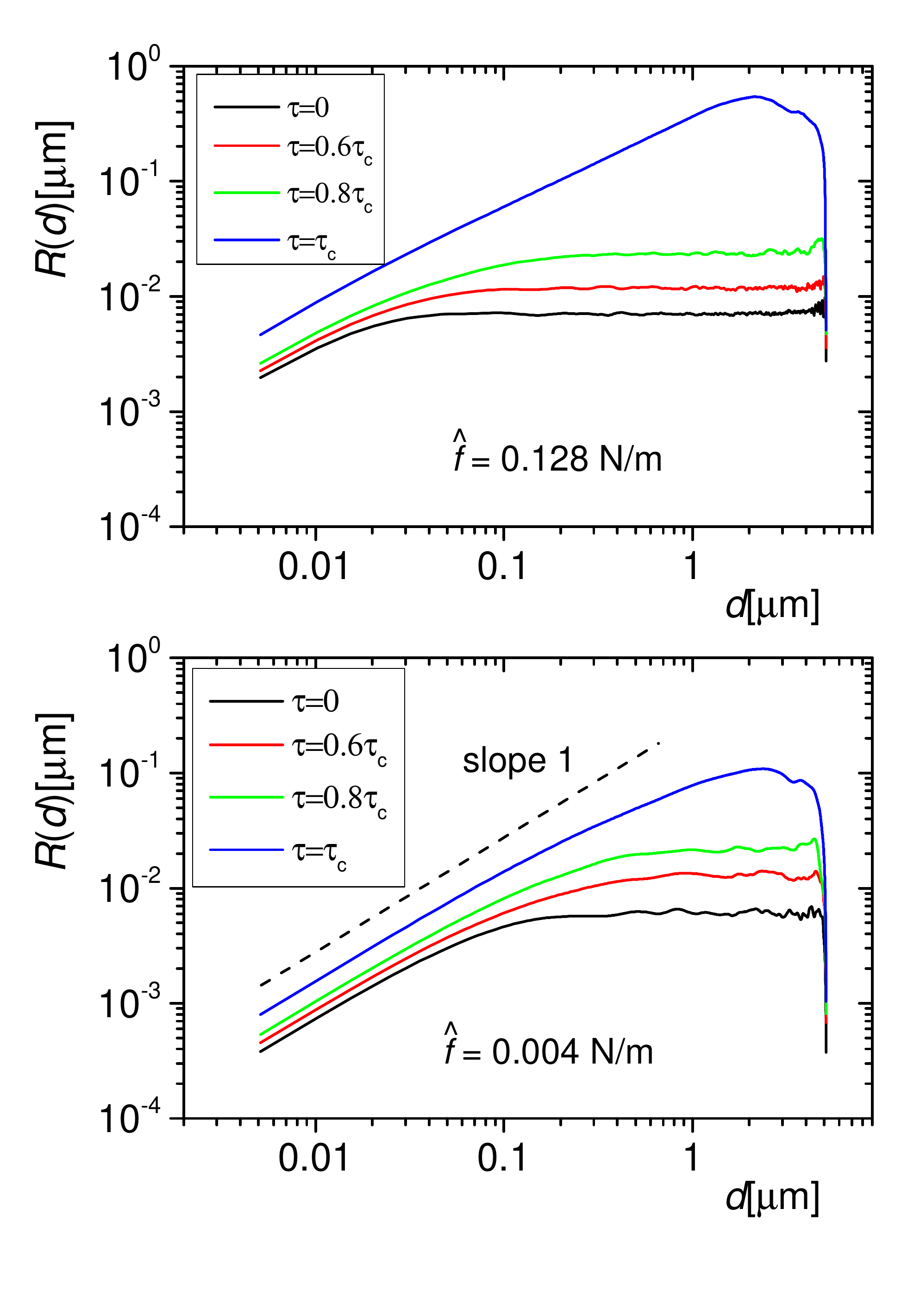}
	\caption{Evolution of the structure function $R(d,\tau)$ under increasing applied shear stress in a stationary pinning field; top:
	amplitude of the pinning field $\hat{f}$ = 0.128 N/m, bottom: $\hat{f}$ = 0.004 N/m; the resolved shear stress is given as a fraction of the critical resolved shear stress (crss), see Figure \ref{fig:crss}, top, for the respective crss values.\label{fig:roughquenched}}
\end{figure}

\begin{figure}[hbt]
	\centering
	\includegraphics[width=3.3in]{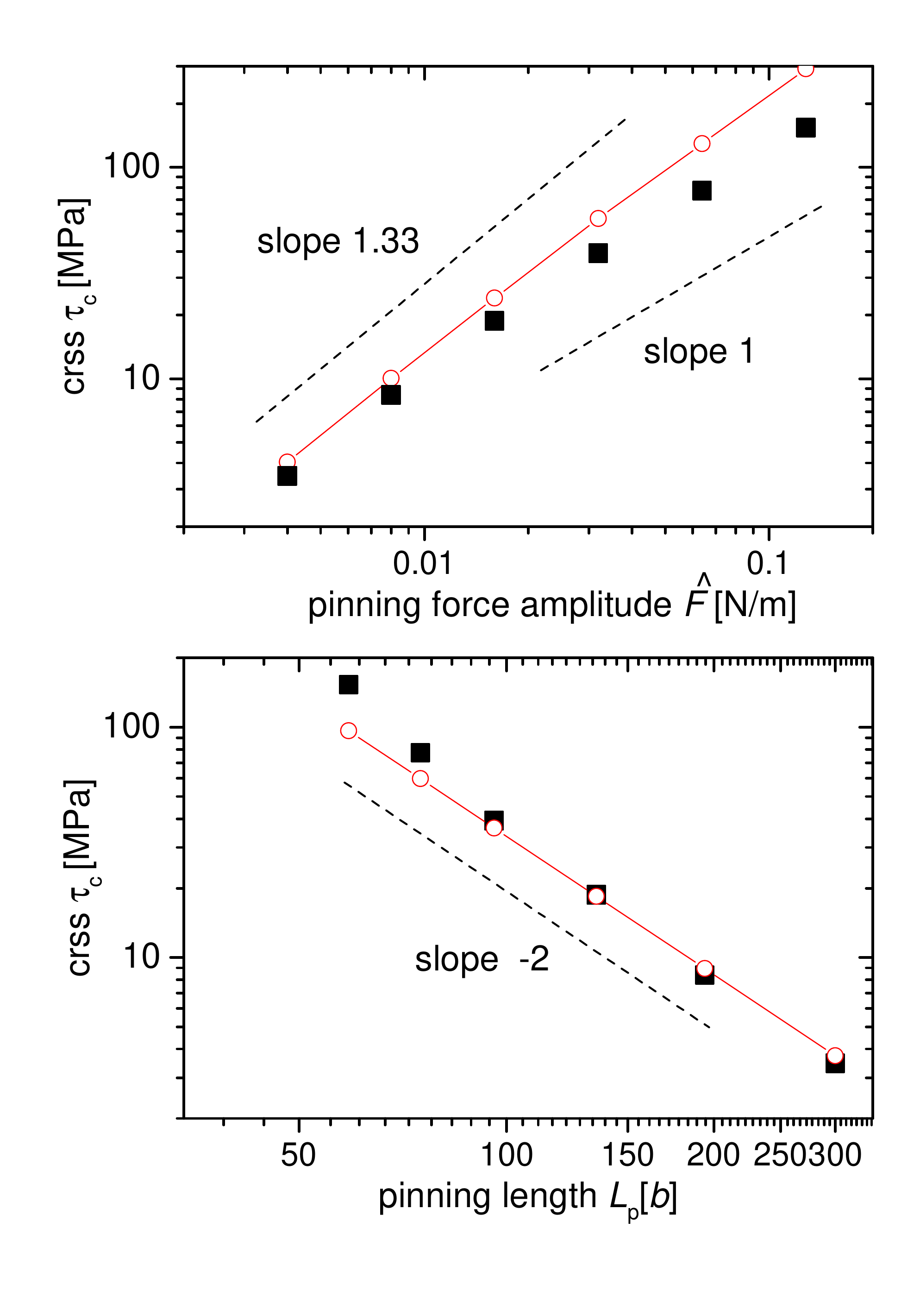}
	\caption{Top: critical resolved shear stress as function of amplitude $\hat{f}$ of the random force field, full squares: crss values
	determined from our simulations, open circles: 	prediction from Eq. (\ref{eq:fpin}); bottom: critical resolved shear stress as function of pinning length (pinning length as in Figure \ref{fig:pinlength}), full squares: crss values determined from our simulations, open circles: 	prediction from Eq. (\ref{eq:fpinL}). \label{fig:crss}}
\end{figure}

We now turn to the behavior of the dislocation line under applied stress. An increasing applied stress causes the dislocation to move between metastable configurations until, at a critical stress, metastability is lost and the dislocation moves indefinitely (depinning). In terms of the self affine roughness of the dislocation line, roughening extends to larger and larger scales until, at the critical stress, the dislocation exhibits a self affine shape on all scales with roughness exponent $\zeta \approx 1$. This roughening is illustrated in Figure \ref{fig:roughquenched} for two different amplitudes $\hat{f}$ of the pinning field. 
We define the critical resolved shear stress (crss) as the minimum stress required to move the dislocation once across the periodic simulation 
cell of width $2000b$ in the (mean) direction of dislocation motion. To compare with the prediction for an elastic line, Eq. (\ref{eq:fpin}),  we need to specify a line energy: We identify the wave vector corresponding to the pinning length as $q_{\rm p} = 2 \pi/L_{\rm p}$ and obtain the corresponding line energy from Figure \ref{fig:eqdis}, thus assuming that pinning is controlled by an effective value of the line energy on the scale of the pinning length. This allows us to calculate a pinning force from  Eq. (\ref{eq:fpin}), from which the predicted crss derives by simply setting $f_{\rm c} = \tau_{\rm c}b$. Predicted crss values are compared to values deduced from the simulations in Figure \ref{fig:crss}, top. It can be seen that in the regime of weak pinning, here up to a force amplitude of about $\hat{f}=$0.02 N/m, corresponding to a pinning length of about 100$b$, the simulated crss data (black squares in Figure \ref{fig:crss}, top) follow to a good approximation the prediction for an elastic line (red circles in Figure \ref{fig:crss}, top). Deviations occur in the regime of strong pinning, corresponding to the regime where the pinning length deviates from the weak pinning result and tends towards a saturation: In this regime, we see a cross-over from an exponent $p=4/3$ in the $\tau_{\rm c} \propto \hat{f}^p$ relationship to an exponent of 1. The transition occurs at a pinning length of 
$L_{\rm p} \approx 5 \xi$, for larger pinning lengths the weak pinning relations are well fulfilled.

We next turn to the possibility of predicting the crss on the basis of the pinning length alone using Eq. (\ref{eq:fpin}). As can be seen in Figure \ref{fig:crss}, bottom, this equation produces excellent results in the weak pinning regime $L_{\rm p} > 100 b = 5 \xi$. 
This finding is important because it suggests that, in the weak pinning regime, estimates of the zero-temperature critical resolved shear stress can be obtained by studying the static properties of a dislocation at zero stress on scales of the pinning length only - scales that are accessible to molecular or even ab-initio simulation methods. Depinning, on the other hand, involves complex spatio-temporal processes and avalanche dynamics covering a wide range of spatial and temporal scales, which require mesoscopic methods such as discrete dislocation dynamics for their simulation.

\section{Discussion and Conclusions}
\label{sec:5}

We have analyzed the roughening of a dislocation line under the influence of thermal forces which has provided us with a method to determine, 
for the DDD model analyzed, a length-scale dependent effective line energy that controls the energy associated with perturbations of a straight line. We then studied the behavior of the simulated dislocation in a stationary random field created by pinning centers of spacing and interaction range $\xi$. Using scaling relations typical of elastic lines in the regime of weak pinning, we found that we can determine a pinning length $L_{\rm p}$ from analyzing the relaxed shape of a initially straight dislocation at zero applied stress. The main result of our investigation is that, in the weak pinning regime where $L_{\rm p} > \xi$, it is possible to obtain estimates of the depinning stress (the critical resolved shear stress needed to move the dislocation across the pinning field) without explicit knowledge of the pinning field strength, based upon the pinning length, the interaction range $\xi$ which relates to the density of the pinning obstacles, and the scale dependent dislocation line energy. 

What is the use of this result? First of all we note that determining the pinning length can be done by static relaxation of the dislocation in a comparatively small periodic simulation cell. For the construction shown in Figure \ref{} to work, it is sufficient that the simulation cell extends over about 5 pinning lengths in the direction parallel to the dislocation and about 5 $\xi$ in the perpendicular direction. Depinning, on the other hand, is preceded by a sequence of increasingly complex metastable configuurations where the dislocation develops self affine roughness on all scales and, as can be inferred from Figure \ref{}, accurate determination of the flow stress requires a simulation cell that extends over about $1000 \xi$ in the direction parallel and over $100\xi$ in the direction perpendicular to the dislocation. Hence, the size of the simulation cells needed for determination of $\tau_{\rm c}$ and of $L_{\rm p}$ differs by a factor of about 200. In addition, direct determination of the crss needs a search algorithm to find the critical stress at which, in such a simulation, metastability is lost, which necessitates a sequence of relaxation steps or a simulation of the time dependent dynamics under a very slowly ramping external stress. 

In a conventional solution hardened alloy, e.g. a binary alloy system for which reliable inter-atomic potentials are available, it is nevertheless possible to perform medium-scale molecular dynamics simulations to directly determine the crss \cite{Rodary2004_PRB} from atomic simulation. This direct approach cuts out the need to parametrize a meso-scale model: In a concentrated solid solution the solute spacing and hence the correlation length $\xi$ is expected to be of the order of $\xi \sim 2b$ only, hence a system of size $100 \times 100 \times 1000 \xi^3$ as used in our study is accessible to large-scale MD simulations. In medium- or high-entropy alloy systems for which phenomenological potentials are available, similar simulations may be performed as demonstrated by \cite{Rao2017_AM} for the quarternary CoFeNiTi system. The here presented method may then allow for a computationally efficient screening of the space of compositions within such an alloy system in view of establishing optimal mechanical properties, by performing serial simulations of small systems and establishing the composition dependence of the pinning length from the respective relaxed dislocation configurations. 

To show that our ideas are indeed applicable to high-entropy alloys we refer to experimental data from the literature. For fcc equi-atomic CoCrFeMnNi alloy a solute contribution to strength of about 300 MPa has been reported \cite{Otto2013_AM} while the shear modulus amounts to $\mu$=81 GPa and the Young's modulus to 203 MPa \cite{Laplanche2015_JAC}. Using a line energy estimate of $0.3 \mu b^2$ and correlation length $\xi \approx b$ we find from Eq. (\ref{eq:fpin}) a characteristic pinning length $L_{\rm p} > 9 \xi$ which indicates that this alloy is well within the weak pinning regime (see also the work of Varvenne et. al. \cite{Varvenne2016_AM} who apply weak-pinning type averaging methods to the same alloy). For bcc equi-atomic TaNbHfZrTi the solute contribution to strength is about 700MPa \cite{Senkov2011_JAC} which with an estimated shear modulus of $\mu$=40 GPa \cite{Senkov2011_JAC} and similar line energy and correlation length estimates implies a characteristic pinning length $L_{\rm p} \approx 5 \xi$, still at the borderline of the weak pinning regime. Finally, in a computational study of a quarternary CoFeNiTi alloy data determined by \cite{Rao2017_AM} inidcate a disorder contribution to the $T=0$ crss of screw dislocations of $0.02 \mu$ (difference between the crss values of the random alloy and the disorder-homogeneized alloy), which corresponds to a disorder-associated pinning length of $L_{\rm p} \approx 7 \xi$, again within the weak pinning regime. It is thus likely that weak pinning ideas can be applied to HEA or, more generally speaking, to compositionally complex random alloys, and that the method proposed here allows for a rapid screening of the composition space in such alloy systems where reliable potentials are available. We note that, in fcc HEA, an additional complication arises from the typically low stacking fault energy and resulting high degree of core splitting, which implies that the concept of solute pinning must be applied to split dislocations of even independently pinned particals. We will discuss the resulting modifications to the present considerations elsewhere but note that, as long as the splitting distance is larger than the pinning length, the analysis of weak pinning can be applied to the two partial dislocations separately.

\section*{Acknowledgements}
\addcontentsline{toc}{section}{Acknowledgements}
This work is funded by Deutsche Forschungsgemeinschaft (DFG) under grant 1Za-8/1. Discussions with Daniel Weygand and Volker Mohles are gratefully acknowledged.
\appendix

\section{Derivation of Langevin force on a dislocation}
\label{app:1}
Without external applied stress and spatial force field Eq. (\ref{equ:1}) can be rewritten as
\begin{equation}
\label{equ:a1}
m_{0}L_{i} \bm{a}_{i}=-BL_{i} \bm{v}_{i} + \bm{F}_{i}^{\mathit{disloc}} + \bm{F}_{i,T} 
\end{equation}
where $\bm{F}_{i}^{\rm{disloc}}=-U_{\bm r_{i}}$, $U$ is total potential energy of the dislocation network \cite{Bulatov2006_OUP} and $\bm r_{i}$ is the position of node $i$. $L_{i}$ is the effective segmeint length associated with node $i$ and can be expressed as:
\begin{equation}
L_{i} = \sum\limits_{j=1}^N0.5L_{ij} 
\end{equation}
where $ j $ labels the $N$ nodes connected to node $ i $ by connecting segments of length $L_{ij}$. For a straight line not connected into 
a network, as considered here, $N =2$. Eq. (\ref{equ:a1}) can be rewritten as following:
\begin{equation}
\bm{a}_{i} + \gamma\bm{v}_{i} = \bm{\mathit{\Gamma}}_{i}^{\rm disloc} + \bm{\mathit{\Gamma}}_{i,T}
\end{equation}
where $\gamma=B/m_{0}$ and $\bm{\mathit{\Gamma}}_{i}=\bm{\mathit{F}}_{i}/(m_{0}L_{i})$ denotes a force per unit mass.

The Langevin force describing a thermal white noise has the following statistical properties \cite{Risken1996_SPRINGER}:
\begin{equation}
\begin{split}
&\left\langle \mathit{\Gamma}_{i,T}\left( t\right)\right\rangle = 0  \\
&\left\langle {\mathit{\Gamma}}_{i,T}(t){\mathit{\Gamma}}_{i,T}(t')\right\rangle  = q\delta(t-t')
\end{split}
\label{equ:a4}
\end{equation}
where the noise strength $q$ is given by
\begin{equation}
q=2\gamma k_{B}T/(m_{0}L_{i})=2Bk_{B}T/(m_{0}^{2}L_{i})
\end{equation}
where $k_{B}$ is Boltzmann constant and $T$ is absolute temperature. In a numerical implementation, we statistically integrate the 
Langevin force over the elementary time step, yielding 
\begin{equation}
{\mathit{\Gamma}}_{i,T}(t_{j} \leq t<t_{j}+\Delta t_{j})=\Lambda_{i,j}\sqrt{\frac{2Bk_{B}T}{m_{0}^{2}L_{i}\Delta t_{j}}}
\end{equation}
where $\Lambda_{i,j}$ is a Gaussian distributed random variable of zero mean and unit variance, and $\Delta t_{j}$ is the length of simulation time step $j$. Accordingly, the Langevin force on a dislocation segment of length $L_{i}$ has the following form:
\begin{equation}
\label{equ:a7}
{\mathit{F}}_{i,T}( t_{j} \leq t<t_{j}+\Delta t_{j})  = \Lambda_{i,j}\sqrt{\frac{2Bk_{B}TL_{i}}{\Delta t_{j}}}
\end{equation}
and the Langevin force per unit length is:
\begin{equation}
\label{equ:a8}
{\mathit{f}}_{i,T}(t_{j} \leq t<t_{j}+\Delta t_{j})=\Lambda_{i,j}\sqrt{\frac{2Bk_{B}T}{L_{i}\Delta t_{j}}}
\end{equation}

Following the suggestion of \cite{Mohles1996_TUB}, the component of the Langevin force along the dislocation line direction is ignored to avoid numerical problems and the nodal Langevin force is thus given by the following expression:
\begin{equation}
\bm{s}_{i} = \sum\limits_{j=1}^NL_{ij}\bm{s}_{ij}/|\sum\limits_{j=1}^NL_{ij}\bm{s}_{ij}|
\end{equation}
where $\bm{s}_{ij} = \bm{n}_{ij}\times\bm{t}_{ij}$ is the glide direction of segment $i-j$, $\bm{n}_{ij}$ and $\bm{t}_{ij}$ are the unit normal vectors of glide plane and the unit line direction vector of dislocation segment $i-j$, respectively.
Combining magnitude and direction gives the Langevin force and force per unit length as:
\begin{equation}
\begin{split}
&\bm{\mathit{F}}_{i,T} = {\mathit{F}}_{i,T}\bm{s}_{i} \\
&\bm{\mathit{f}}_{i,T} = {\mathit{f}}_{i,T}\bm{s}_{i}
\end{split}
\end{equation}


\section*{References}
\addcontentsline{toc}{section}{References}
\bibliographystyle{elsarticle-num} 
\bibliography{references}






\end{document}